\documentclass[usenatbib]{mnras}
\usepackage{graphicx,times}
\usepackage{txfonts}
\usepackage{color}

\title[Jellyfish galaxies in galaxy cluster mergers]{Galaxy cluster mergers as triggers for the formation of jellyfish galaxies: case study of the A901/2 system}
\author[Ruggiero et al.]{Rafael Ruggiero$^{1}$\thanks{E-mail: rafael.ruggiero at usp.br}, Rubens E.~G.~Machado$^{2}$, Fernanda V. Roman-Oliveira$^{3}$, \and Ana L. Chies-Santos$^{3}$, Gast\~ao B.~Lima Neto$^{1}$, Lia Doubrawa$^{2}$, Bruno Rodr\'iguez del Pino$^{4}$
\\
$^{1}$Instituto de Astronomia, Geof\'isica e Ci\^encias Atmosf\'ericas, Universidade de S\~ao Paulo, R. do Mat\~ao 1226, 05508-090 S\~ao Paulo, Brazil\\
$^{2}$Departamento Acad\^emico de F\'isica, Universidade Tecnol\'ogica Federal do Paran\'a, Rua Sete de Setembro 3165, Curitiba, Brazil\\
$^{3}$Departamento de Astronomia, Instituto de F\'isica, Universidade Federal do Rio Grande do Sul, Porto Alegre, Brazil\\
$^{4}$Centro de Astrobiolog\'ia, CSIC-INTA, Torrej\'on de Ardoz, 28850, Madrid, Spain\\
}
\date{Accepted 0000 000 00. Received 0000 000 00; in original form 0000 000 00}
\pubyear{2018}

\begin{document}  
\label{firstpage}
\pagerange{\pageref{firstpage}--\pageref{lastpage}}
\maketitle

\begin{abstract}
The A901/2 system is a rare case of galaxy cluster interaction, in which two galaxy clusters and two smaller groups are seen in route of collision with each other simultaneously. Within each of the four substructures, several galaxies with features indicative of jellyfish morphologies have been observed. In this paper, we propose a hydrodynamic model for the merger as a whole, compatible with its diffuse X-ray emission, and correlate the gas properties in this model with the locations of the jellyfish galaxy candidates in the real system. We find that jellyfish galaxies seem to be preferentially located near a boundary inside each subcluster where diffuse gas moving along with the subcluster and diffuse gas from the remainder of the system meet. The velocity change in those boundaries is such that a factor of up to $\sim$1000 increase in the ram pressure takes place within a few hundred kpc, which could trigger the high rate of gas loss necessary for a jellyfish morphology to emerge. A theoretical treatment of ram pressure stripping in the environment of galaxy cluster mergers has not been presented in the literature so far; we propose that this could be a common scenario for the formation of jellyfish morphologies in such systems.
\end{abstract}

% Select between one and six entries from the list of approved keywords.
\begin{keywords}
galaxies: clusters: general -- galaxies: interactions -- methods: numerical
\end{keywords}

%%%%%%%%%%%%%%%%%%%%%%%%%%%%%%%%%%%%%%%%%%%%%%%%%%%
\section{Introduction}

% LambdaCDM, haloes and mergers
In a $\Lambda$CDM cosmology, primordial inhomogeneities in the density field of the Universe are expected to act as seeds for the later formation of structures. On small scales, gravity tends to make initially small inhomogeneities evolve into collapsed structures, most notably dark matter haloes, which later become the hosts of objects such as galaxies and galaxy clusters. In accordance with $\Lambda$CDM, such haloes often interact with each other through mergers; galaxy cluster mergers are the most extreme version of such interactions, and are the most energetic events in the universe since the Big Bang \citep{Sarazin2002}.

% A901/2 system. 
One prominent example of a system in interaction is the A901/2 multi-cluster, at $z \sim 0.165$. This is an unrelaxed system that contains four main cores -- A901a, A901b, A902 and the SW group -- and provides an ideal laboratory for probing galaxy evolution along different scales of environment and galaxy masses (\citealt{gray04}, \citealt{Gray2009}). All four subclusters are at similar redshifts \citep[see e.g.][]{Weinzirl2017}, and the two most massive cores (A901a and A901b) have overlapping virial radii, which indicates that the system is likely a multi-cluster merger in its early stages. This is reinforced by the fact that a system with the mass of A901/2 ($\sim$$3.5\times10^{14}$ M$_\odot$) is expected to collapse if its spacial extent is smaller than about 5 Mpc \citep[][]{busha03}, while most of the mass in A901/2 is within a spacial scale of a few Mpc.

% Past simulations of galaxy cluster mergers
Numerical simulations have often been employed to study mergers of galaxy clusters, both from a more general, theoretical point of view, and also in order to model specific objects. Binary cluster collisions are particularly well suited for this purpose, because the numerical resolution can be entirely focused on the objects of interest, as opposed to fully cosmological simulations of structure formation. For example, the Bullet Cluster has been studied in this way \citep{Springel2007, Mastropietro2008, Lage2014}, as have other so-called dissociative clusters \citep[e.g.][]{Donnert2014, MachadoMonteiro2015, Molnar2015}, in which gas and dark matter are offset as a result of the collision. Numerous other observed clusters have been modelled by dedicated simulations that aim to reconstruct their dynamical histories. Simulations have been used to study several phenomena related to collisions of galaxy clusters, such as radio relics \citep[e.g.][]{vanWeeren2011}, sloshing cold fronts \citep[e.g.][]{ZuHone2010, Machado2015, Walker2018}, turbulence \citep[e.g.][]{ZuHone2013a, Vazza2012}, thermal conduction \citep[e.g.][]{ZuHone2013b}, etc. Tailored simulations involving more than two initial objects are more uncommon. For example, a triple merger has been simulated by \cite{Bruggen2012} in order to model \mbox{1RXS~J0603.3+4214}.

% Ram pressure
Gas-rich galaxies which move within the environment of galaxy clusters are expected to have their evolution affected by the interaction with the intracluster medium (ICM). The ram pressure exerted by the ICM can lead to gas loss by ram pressure stripping \citep{1972ApJ...176....1G}, which, in more extreme cases, leads to the formation of ``jellyfish morphologies'', in which the galaxy is observed featuring a filamentary tail of stripped gas and stars. The phenomenon of ram pressure stripping of cluster galaxies has been extensively modelled through numerical simulations, which have explored e.g. the role of inclination angle in the rate of gas loss \citep{2006MNRAS.369..567R}, the changes in star formation rate which take place within the disks of affected galaxies \citep{2008A&A...481..337K,2012A&A...544A..54S,Ruggiero2017}, and the predicted emission features within their tails \citep{2009A&A...499...87K,2010ApJ...709.1203T}. Although most of the numerical work on ram pressure stripping has been based on idealised setups, cosmological simulations of galaxy formation have also been used to explore the phenomenon, as e.g. in \citet{2007ApJ...671.1434T} and more recently in \citet{2018arXiv181000005Y}. 

% Connecting ram pressure vs galaxy cluster merger
Jellyfish galaxies have been found in large numbers in different cluster systems (see e.g. \citealt{ebeling14}, \citealt{Poggianti16}). However, the number of jellyfish galaxies found in single systems is usually small.
The numbers range from 21 in Coma (\citealt{smith10}, \citealt{yagi10}), 3 in Virgo (\citealt{abramson16}, \citealt{kenney14}, \citealt{kenney99}), 1 in A3627 (\citealt{sun06}) and 5 in A2744 (\citealt{rawle14}).
Nevertheless, the rich population of $\sim$70 jellyfish galaxy candidates found in the A901/2 system (\citealt{roman-oliveira18}) indicates that clusters in interaction may be an ideal environment to search for these galaxies.
Moreover, \cite{2016MNRAS.455.2994M} performs a large systematic search for such jellyfish morphologies and suggests that galaxy cluster mergers are more likely to be triggering extreme ram pressure stripping events. This scenario has also been suggested in \cite{owers12}, where four jellyfish galaxies were found near merger signatures of the gas. It is not surprising that such relation could exist: in galaxy cluster mergers, higher ICM velocities are found than in isolated clusters, making those environments favourable for the formation of jellyfish structures.

% What is new here
In this work, we attempt to probe the physical mechanism behind the formation of jellyfish galaxies in galaxy cluster mergers. For that, we model the diffuse gas in the A901/2 system with a galaxy cluster merger simulation, and then compare the gas conditions in this model to the location of a sample of jellyfish galaxies found in this system, allowing us to infer a scenario for the triggering of jellyfish morphologies both in the A901/2 system and in galaxy cluster mergers in general. Such theoretical treatment of ram pressure stripping in galaxy cluster mergers has not been given so far in the literature.

% Structure of the paper
This paper is structured as follows. In Section \ref{sec:sample}, we describe the sample of jellyfish galaxy candidates we use and comment on how they were selected. Then we proceed to describe the setup and the results of our galaxy cluster merger simulation in Section \ref{sec:simulation}. The gas conditions in this simulation are correlated with the locations of the jellyfish galaxies in our sample in Section \ref{sec:results}, where we tentatively propose a physical mechanism for the generation of many of those jellyfishes. Finally, our results are discussed and summarised in Section \ref{sec:discussion}, where possible extensions of our work are also presented.

%%%%%%%%%%%%%%%%%%%%%%%%%%%%%%%%%%%%%%%%%%%%%%%%%%%
\section{Sample of galaxies} \label{sec:sample}

The galaxies used in this study come from an extensive search for galaxies with jellyfish morphological signatures in the A901/2 system \citep{roman-oliveira18}. This sample was selected through visual inspection of HST/ACS F606W images of galaxies in the parent sample of H$\alpha$ emitting galaxies in the OMEGA survey \citep{chies-santos15,rodriguezdelpino17,Weinzirl2017,wolf18}.
The visual inspection method applied follows the work of \citet{ebeling14} and \citet{Poggianti16}. A classification in JClasses was also employed, in which a number from 1 to 5 is assigned to a galaxy to evaluate its degree of asymmetry -- larger values are correlated with a greater likelihood of the galaxy being an actual jellyfish.

The final sample is restricted to the most reliable cases of jellyfish candidates, which we take as those classified as JClass 3 to 5. This sample contains the 73 galaxies that we use in this work. The image stamps are available at the OMEGA jellyfish candidates ATLAS\footnote{OMEGA jellyfish candidates ATLAS: \url{http://lief.if.ufrgs.br/~fernandavro/atlas.pdf}}.

%%%%%%%%%%%%%%%%%%%%%%%%%%%%%%%%%%%%%%%%%%%%%%%%%%%
\section{Simulations} \label{sec:simulation}

This work is based on a galaxy cluster merger simulation including the dark matter haloes and intracluster gas of the 4 subclusters in the A901/2 system, which was used to reproduce their positions on the plane of the sky, along with their observed X-ray properties. The simulations were run with the code \textsc{gadget-2} \citep{Springel2005}; in Appendix \ref{sec:codecomparison}, we also briefly compare the gas conditions in the main simulation with its results when it is run in \textsc{ramses} \citep{Teyssier2002}, in order to assess its robustness against a change in numerical methodology.

\subsection{Simulating the system as a whole}

% goal
Here we describe the simulation setup, in which the four subclusters were included with the goal of obtaining a suitable model of the system as a whole. Our aim here was chiefly to recover the relative distances between four subclusters having the known virial masses and also having plausible gas content. The main observational constraints are the virial masses of the subclusters, derived from gravitational weak lensing \citep{Heymans2008}. 

% assumptions
The redshifts of the subclusters are close to each other \citep[e.g.][]{Weinzirl2017}, so we assume that they are on the same plane. We further assume, for simplicity, that the trajectories of the four subclusters are on the plane of the sky. Virial equilibrium would require velocity dispersions of roughly 1000\,km/s; we drew random velocities but choosing the signs of the Cartesian coordinates such that the subclusters are all incoming, i.e.\ falling towards the centre of mass. It should be noted that our explicit assumption here is that the subclusters are currently infalling towards their first approach, i.e.\ they have not previously collided.

% backwards
As a preliminary step, we represent each cluster as a point mass having the $M_{200}$ from \cite{Heymans2008}. They are assigned velocities as described above, and position coordinates are known straightforwardly from observations. With this information we perform a simple gravitational $N$-body simulation (via direct summation) inverting the sign of time; i.e.\ we simply calculate the orbits backwards in time, for 5\,Gyr. This exercise provides a good approximation for the $t=0$ of the actual hydrodynamical simulation. 

% cluster initial conditions
In the next step, we set up four actual subclusters including dark matter and gas. The method for generating initial conditions is similar to those used in \cite{Machado2015} or \cite{Ruggiero2017}, for example. The dark matter haloes follow a \cite{Hernquist1990} profile:
\begin{equation} \label{eq:hernquist}
\rho_{\rm h}(r) = \frac{M_{\rm h}}{2 \pi} ~ \frac{r_{\rm h}}{r~(r+r_{\rm h})^{3}} \, ,
\end{equation}
where $M_{\rm h}$ is the total dark matter mass, and $r_{\rm h}$ is a scale length. The gas is represented by a \cite{Dehnen1993} density profile (with gas mass $M_{\rm g}$ and scale lenght $r_{\rm g}$), adopting $\gamma=0$:
\begin{equation}
\rho_{\rm g}(r) = \frac{(3-\gamma)~M_{\rm g}}{4\pi} ~ \frac{r_{\rm g}}{r^{\gamma}(r+r_{\rm g})^{4-\gamma}} \, .
\end{equation}
The requirement of hydrostatic equilibrium determines the gas temperatures. Realisations of these initial conditions are created according to the procedures described in \cite{Machado2013}. The virial masses, virial radii and gas fractions of the initial conditions are given in Table~\ref{tb:ic}.

%--------------------------------------------------------------------
\begin{table}
\caption{Initial conditions of the simulated subclusters. The first column gives the names of the models and the objects they are meant to represent. The second and third columns give the virial mass and virial radius. The fourth column gives the overall gas fraction.}
\label{tb:ic}
\begin{center}
\begin{tabular}{l c c c c c}
\hline
         &   $M_{200}$     & $r_{200}$ &  $f_{\rm gas}$\\ 
         &  (${\rm M}_{\odot}$)  & (kpc) &     \\
\hline
subcluster A  (A901a)   & $1.3 \times 10^{14}$ & 1034 & 0.08 \\
subcluster B (A901b)    & $1.3 \times 10^{14}$ & 1036 & 0.15 \\
subcluster C (A902)     & $0.4 \times 10^{14}$ & 688  & 0.08 \\
subcluster D (SW Group) & $0.6 \times 10^{14}$ & 788  & 0.06 \\  
\hline
\end{tabular}
\end{center}
\end{table}
%--------------------------------------------------------------------

% resolution
Each of the four subclusters has $10^6$ gas particles and $10^5$ dark matter particles. Tests of the present simulations indicated convergence across three orders of magnitude in particle numbers, as far as the orbits are concerned. Moreover, since the subclusters are not interpenetrating, their gravitational potentials remain sufficiently spherical in the current stage of the approach. In this specific configuration, one could even attempt to model them by rigid analytic spherical potentials without much loss of detail. We opted to represent them as $N$-body particles. Here we employ the smoothed particle hydrodynamics (SPH) $N$-body code {\sc gadget-2} \citep{Springel2005}, and the evolution is followed for 5\,Gyr.

% fine tuning
The four subclusters, created in the manner described above, are then placed at the locations that were reached by the end of the backwards point-mass integration. And then the subclusters are allowed to evolve forward in time for 5\,Gyr. They fall towards the centre of mass until the current observed separations are reached. However, they do not reach exactly the desired coordinates by the end, because the orbits of four point masses are not identical to the orbits of four extended objects. Some fine tuning of their initial positions and velocities was performed by trial and error until an acceptable agreement was reached. In the resulting preferred model, the instant when the coordinates best matched the observations was $t=4.3$\,Gyr.

%--------------------------------------------------------------------
\begin{figure}
\includegraphics[width=\columnwidth]{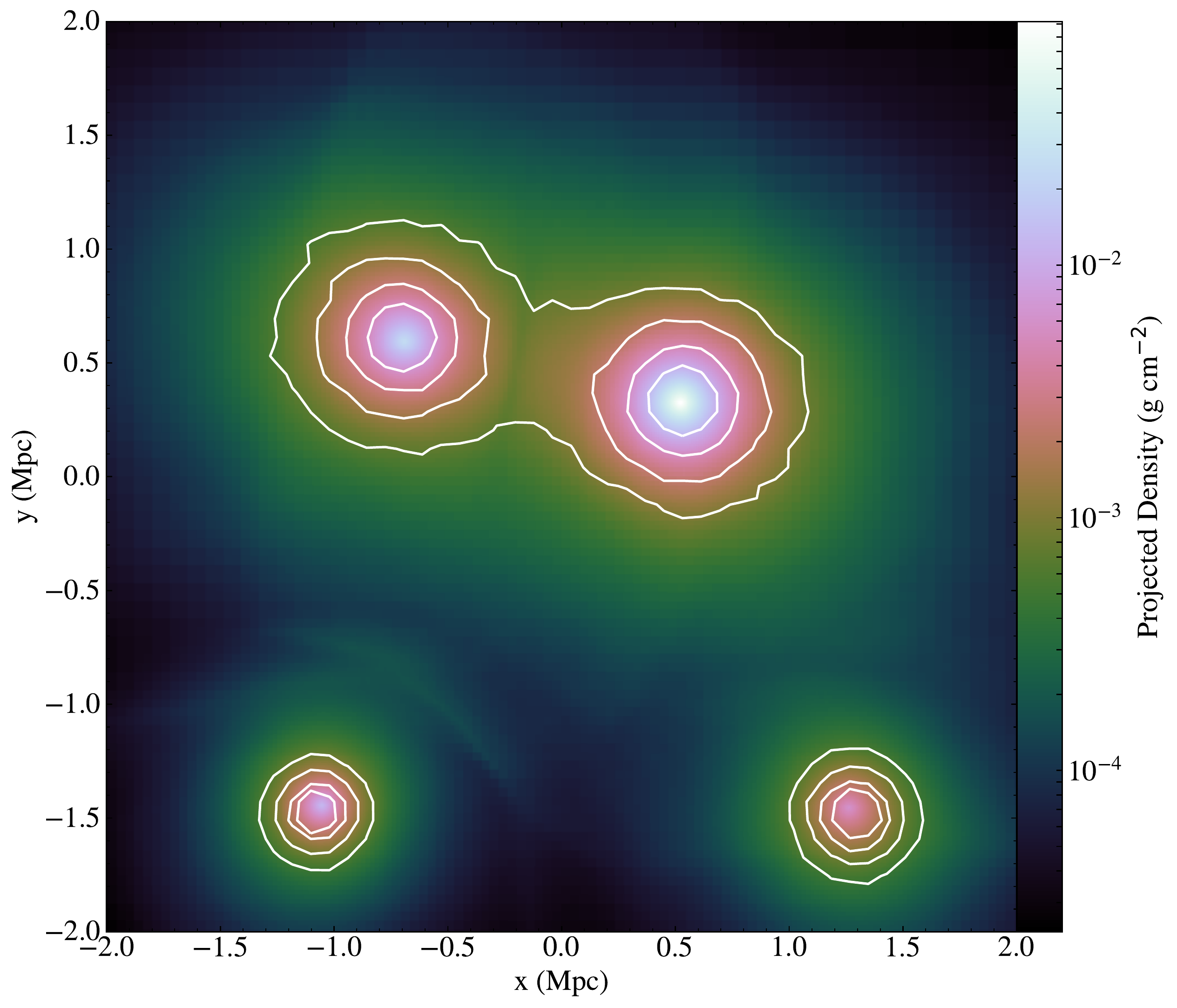}
\caption[]{This is the snapshot that best reproduces the observed relative separations between the subclusters ($t=4.3$\,Gyr). Colours represent the projected gas density. Total projected mass is shown as contours.}
\label{fig:density}
\end{figure}
%--------------------------------------------------------------------

\subsection{X-ray mock image}

% X-rays
Observations indicate that A901b is the only one of the four with significant X-ray diffuse emission. A901a hosts a very bright AGN, so its extended emission is unclear. A902 is barely above the background noise, and the SW Group is essentially undetectable in X-rays \citep{Gilmour2007}. To ensure a higher X-ray emission, subcluster B in Table~\ref{tb:ic} has the highest gas content of the four. In the absence of detailed observational constraints, the other simulated subclusters were chosen to have a low gas fraction of approximately 8 per cent (or 6 per cent in the case of the SW group), towards the lower limit of what is expected for their masses \citep{Lagana2013}. The simulated gas densities of the best-matching instant are shown in Fig.~\ref{fig:density}. In this figure, one may also notice that the centroids of the projected total mass distributions of the four subhaloes reproduce the observed relative separations in a good approximation.

\begin{figure*}
\includegraphics[width=\textwidth]{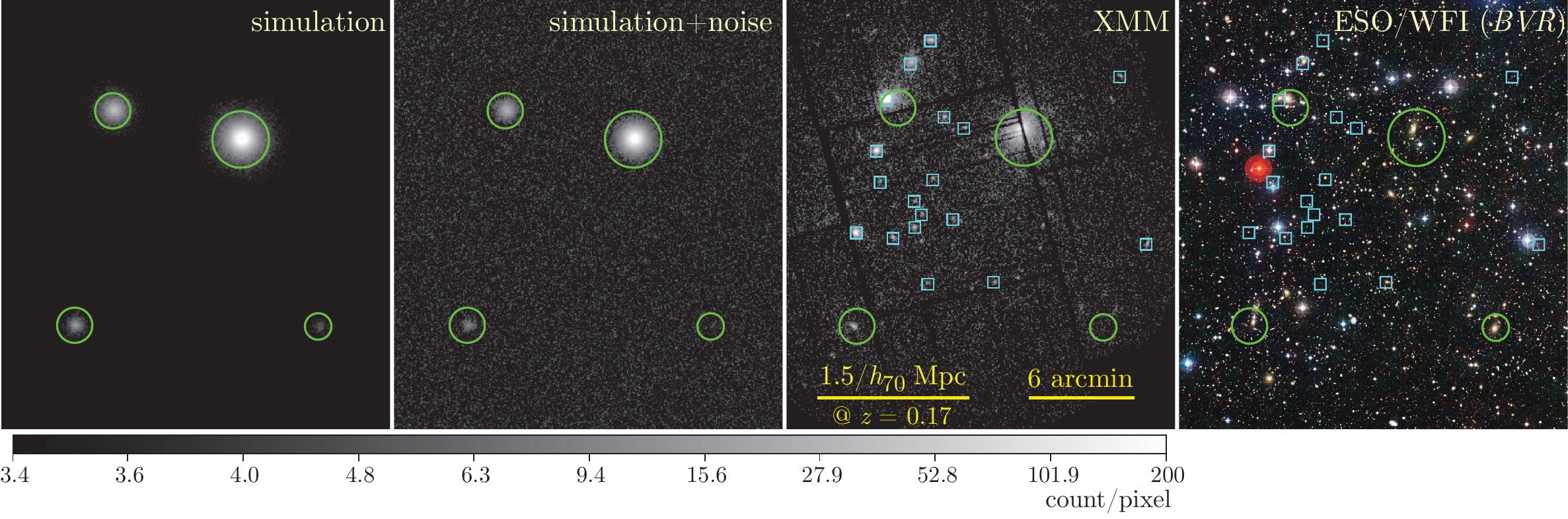}
\caption[]{Comparison between the the mock X-ray image and the observations. First frame: mock X-ray image produced from the simulation. Second frame: same as before, but with added noise. Third frame: \textit{XMM} observation. Fourth frame: Optical image from ESO/WFI.}
\label{fig:A901mockXMM}
\end{figure*}

% mock packages
We performed a more quantitative test to ensure that the simulated gas densities were not excessive. Using the $t=4.3$\,Gyr snapshot of the simulation, we produced a mock X-ray image with the following procedure, assuming thermal emission from a hot plasma. We used \texttt{pyXSIM}\footnote{\url{http://hea-www.cfa.harvard.edu/~jzuhone/pyxsim/}}, a Python package for simulating X-ray observations from astrophysical sources. It is based on an algorithm of \cite{Biffi2012,Biffi2013}, but see also \cite{ZuHone2014}. In brief, it takes as input the simulated densities and temperatures of the gas, assumes a constant metallicity of 0.3 Z$_{\odot}$, and generates a photon sample assuming a spectral model ({\sc apec} from the AtomDB database\footnote{\url{http://www.atomdb.org/}}). The photon sample is then projected along the line of sight (the $z$ axis of the simulation). Given the coordinates of the cluster, a foreground Galactic absorption model is also applied, assuming a neutral hydrogen column of $N_{\rm H} = 4 \times 10^{20}~{\rm cm}^{-2}$. The photon list is exported to be used by the SIXTE\footnote{\url{http://www.sternwarte.uni-erlangen.de/research/sixte/}} (Simulation of X-ray Telescopes) package, to be convolved with the \textit{XMM} instrument response (the EPIC MOS camera, in this case). The effective exposure time was 67\,ks and the energy range was 0.2--7.0\,keV. Poissonian noise was added to the resulting 600$\times$600-pixel mock image, shown in Fig.~\ref{fig:A901mockXMM}. Note that the X-ray emission of the SW group is present in the first frame, albeit very faint. Once noise is added, it is lost in the background.

% good agreement and caveats
Our resulting model is approximate, and it cannot be expected to account for all details of the observed systems. Furthermore, there are no assurances that the solution we have found for the orbits is unique, as is always the case in such reconstructions. However, the gas properties in the model are physically well-motivated, and compatible with the observational expectation of X-ray detections -- two subclusters with significant emission are obtained, plus two near the threshold of detection (bearing in mind that the diffuse emission of A901a is somewhat inconclusive due to the very bright point source). Therefore, the resulting snapshot of the simulation should offer a sufficiently realistic environment in which to study ram pressure effects.

%--------------------------------------------------------------------
%--------------------------------------------------------------------

%%%%%%%%%%%%%%%%%%%%%%%%%%%%%%%%%%%%%%%%%%%%%%%%%%%
\section{Local conditions of the jellyfish galaxies} \label{sec:results}

Now we turn to the analysis of the gas properties in the merger model presented in the previous section, with the goal of answering the question: what explains the presence of jellyfish galaxies at the locations where they are found in A901/2? Naturally, the two most important quantities to be analysed should be the diffuse gas density and the diffuse gas velocity across the system, since jellyfish morphologies are caused by ram pressure stripping events, while the ram pressure $P_{\mathrm{ram}}$ depends on those two quantities \citep{1972ApJ...176....1G}:
\begin{equation} \label{eq:rampressure}
P_{\mathrm{ram}} = \rho_{\mathrm{ICM}} v_{\mathrm{ICM}}^2,
\end{equation}
where $v_{\mathrm{ICM}}$ is the ICM velocity relative to a galaxy under consideration. In generating plots involving those quantities, we have used the Python package {\sc yt} \citep{Turk2011} to deposit the simulation particles into a space-filling grid with a ``cell-in-cloud'' approach.

In our analysis, we focus on the ram pressure calculated in the reference frame of each subcluster, as a first approximation for the ram pressure experienced by its member galaxies. The effect of peculiar velocities of member galaxies relative to their parent cluster will be discussed later. These reference frames are defined by the average speed of the dark matter particles within $r_{200}/3$ of the centre of a given subcluster, with the centre location defined as that of the density peak of this cluster's ICM. We have verified that the results that follow are not sensitive to the choice of the inner radius in the velocity calculation -- using the velocities within radii closer to $r_{200}$ yield similar results, but we find it more meaningful to restrict ourselves to the inner region of each subcluster since that region is in principle less disturbed by tidal effects.

\begin{figure*}
\includegraphics[width=\textwidth]{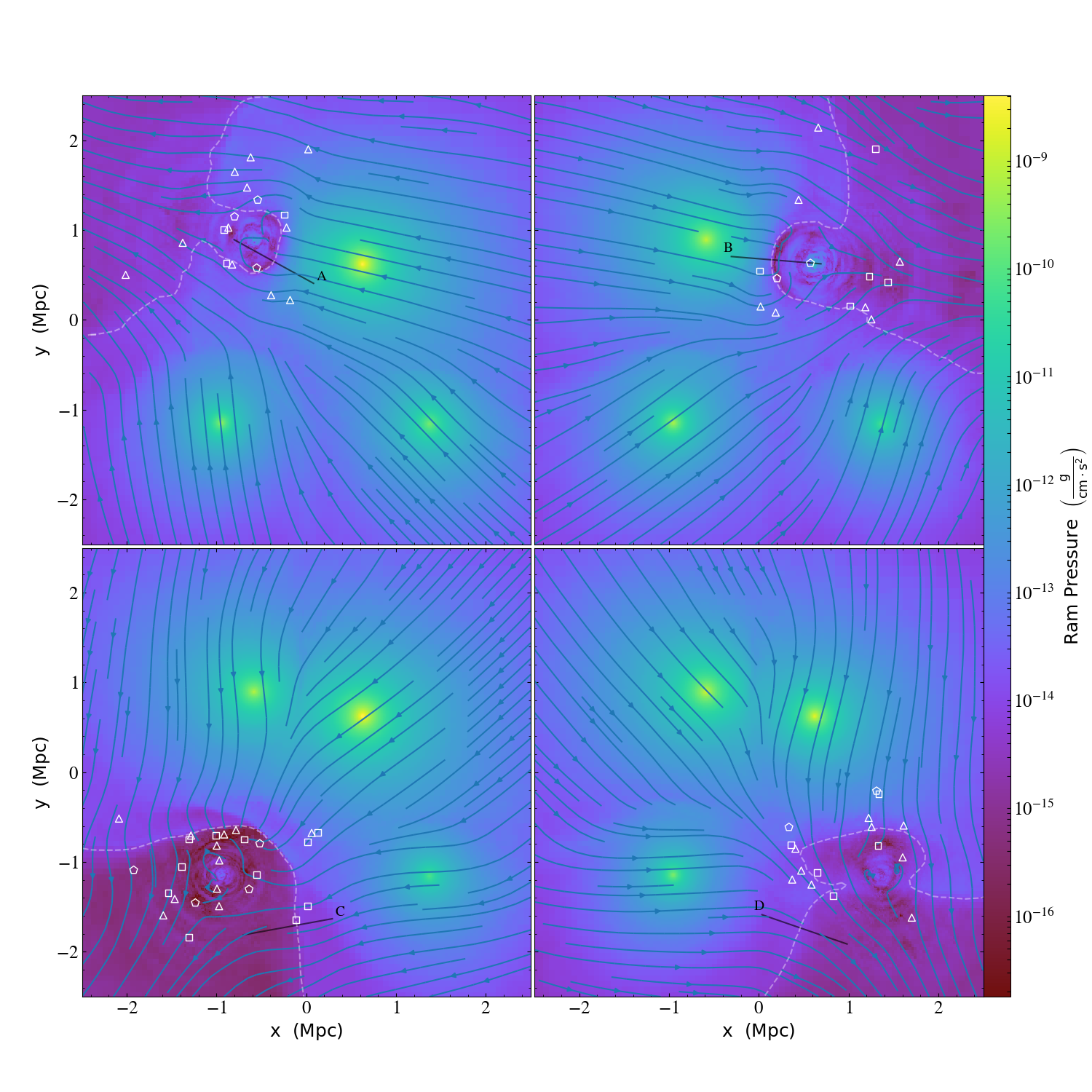}
\caption[]{Ram pressure intensities in the reference frames of each of the four subclusters. Each subplot is a midplane slice. The dashed lines show the approximate locations of the ram pressure boundaries in each subcluster, and triangles, squares and pentagons represent the locations of jellyfish galaxies classified as JClass 3, 4 and 5, respectively. Ram pressure profiles along the four black lines are shown in Fig.~\ref{fig:discontinuity}.}
\label{fig:rampressures}
\end{figure*}

With those four reference frames defined, we are then able to calculate the ram pressure of the system as a whole in each of them. This is shown in Fig.~\ref{fig:rampressures}, where the ram pressure is shown in slices along the plane of the four subclusters in our model, overlaid with streamlines of diffuse gas velocity. At the centre of each subcluster the ram pressure is low, since in that region the diffuse gas is on average moving along with the cluster halo. On the other hand, the ram pressure is intense far from the cluster centre, since the diffuse gas from other clusters is moving in the opposite direction at high speed. It turns out that a reasonably narrow ($\sim$100 kpc) boundary exists between those two regions, where a significant increase in the ram pressure takes place. The dashed contours in Fig.~\ref{fig:rampressures} are the approximate locations of those boundaries, which were obtained using ram pressure isocontours, and in each subplot the positions of the galaxies in our sample closest (in projected space) to the subcluster considered in that plot than to any of the other three are shown.

\begin{figure*}
\includegraphics[width=\textwidth]{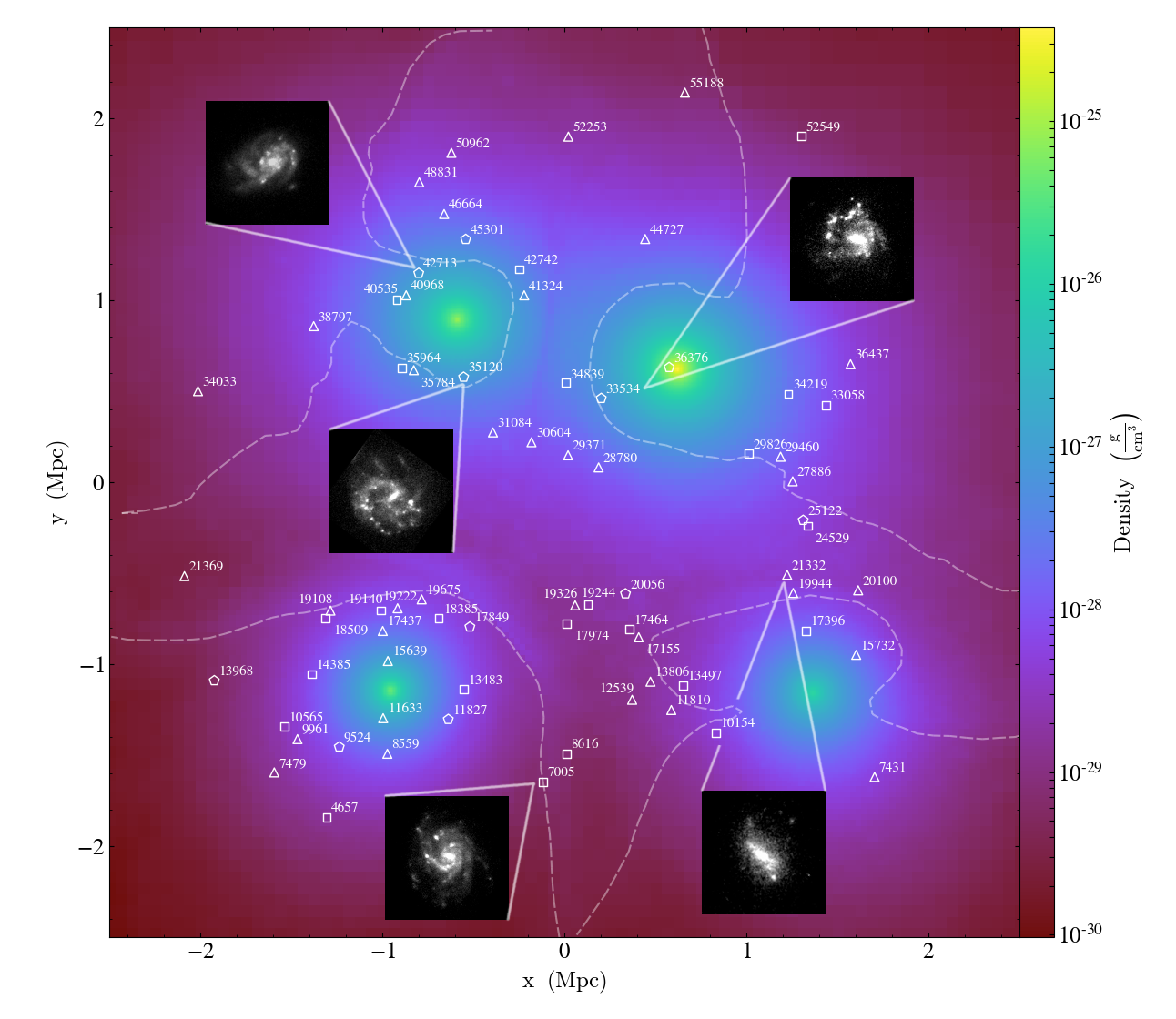}
\caption[]{Identification of our sample of galaxies over a midplane density slice of our simulation. The dashed lines are the locations of the ram pressure boundaries shown in Fig.~\ref{fig:rampressures}. Some HST thumbs in arbitrary scale are shown for the sake of illustration, and, also as in Fig.~\ref{fig:rampressures}, triangles, squares and pentagons are used for JClass 3, 4 and 5 galaxies respectively.}
\label{fig:id}
\end{figure*}

Figure \ref{fig:id} shows a gas density slice of the simulation, in which all the ram pressure boundaries are shown simultaneously, along with the locations of all our jellyfish candidates and some examples of HST images for those galaxies. This plot shows that the density in the system does not feature any pronounced structure at the locations of the boundaries, which implies that they emerge exclusively due to the velocity structure of the diffuse gas around their locations. It is not surprising that this should be the case, since the clusters are approaching each other (and thus their diffuse gas is moving at opposing directions), while the ram pressure depends very strongly on the diffuse gas speed, more so than on its density (see Eq.~\ref{eq:rampressure}). In this way, the ram pressure boundaries can be identified as regions where gas moving along each subcluster and gas from the remainder of the system meet.

%--------------------------------------------------------------------
%--------------------------------------------------------------------

% A901a, A902b, A902, SW group

\begin{figure}
\includegraphics[width=\columnwidth]{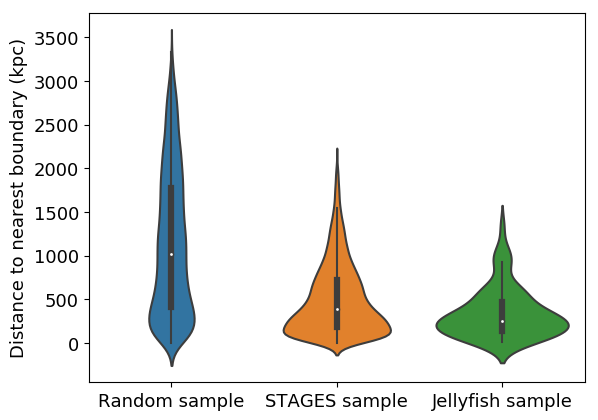}
\caption[]{Violin plot showing the distributions of distances to the nearest ram pressure boundary for a random set of points, the STAGES sample of galaxies in the A901/2 cluster, and our sample of jellyfishes. The jellyfishes are systematically closer to the boundaries than it would be expected if their locations were random, and they are also systematically closer than the non-jellyfish galaxies in the system.}
\label{fig:randomdistances}
\end{figure}

It can be visually noted in Fig.~\ref{fig:rampressures} that many galaxies are located in the vicinities of the ram pressure boundaries. This is not always the case -- for instance, many galaxies in the A902 subcluster are found in a region without an apparent connection to the boundary we find for that subcluster. Still, a correlation seems to exist. We quantify this effect in the following manner. First, we measure the projected distance from each jellyfish galaxy to its respective nearest boundary. Then, we generate a random cloud of points occupying the same area as those galaxies and also measure their distances to the nearest boundaries. The comparison between these two distributions of distances is shown in Fig.~\ref{fig:randomdistances}, which makes it evident that the jellyfishes are systematically closer to a boundary than what would be expected from a random distribution. The Kolmogorov-Smirnov test applied to cumulative, normalised histograms for both samples indicates that the chance of the two distributions being equivalent is very small, of 1 in $\sim$85 million (p-value of $10^{-6}$ per cent). We also make the same comparison using the STAGES sample \citep{Gray2009} of galaxies in the A901/2 cluster instead of a random sample, filtered for member galaxies with stellar mass between $10^{8}$\,M$_\odot$ and $10^{12}$\,M$_\odot$. The upper quartile, lower quartile and median for the jellyfish distribution are all lower than for the STAGES distribution, with a chance of 1 in 757 (p-value of 0.13 per cent) of the two distributions being equivalent, further reinforcing our thesis that the jellyfishes are systematically closer to the ram pressure boundaries we report.

%--------------------------------------------------------------------
%--------------------------------------------------------------------

\begin{figure}
\includegraphics[width=\columnwidth]{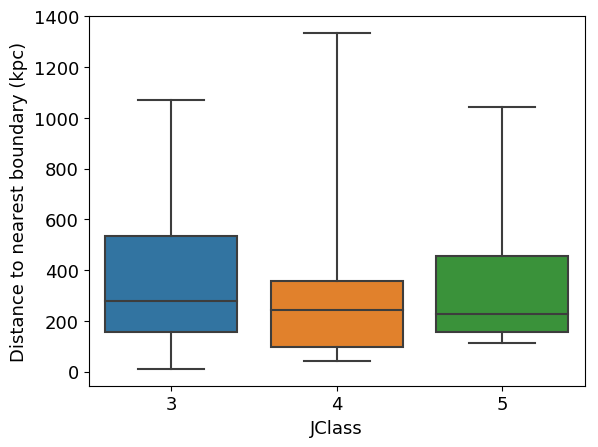}
\caption[]{Box plot showing the distances between the jellyfish galaxies and the nearest ram pressure boundaries as a function of JClass. The median distances decrease with JClass: they are 279 kpc, 245 kpc and 226 kpc for JClass 3, 4 and 5 galaxies, but a Kolmogorov-Smirnov test applied to the pairs of distributions lead us to conclude that this is not a statistically significant result.}
\label{fig:distances}
\end{figure}

The distribution of distances to the nearest boundary can also be analysed as a function of JClass. This is shown in Fig.~\ref{fig:distances}. The three distributions are overall quite similar, but an interesting feature is that the median distance to the nearest boundary decreases systematically with JClass. This could be an indication that jellyfish morphologies are more pronounced when a galaxy has just encountered a boundary, and then on a short timespan after that, they become less intense. The median distance for JClass 5 galaxies is 53\,kpc smaller than for JClass 3 galaxies; assuming that the galaxies move at 1000\,km/s, this would imply that the transition from JClass 5 to 3 happens on a timescale of 53\,Myr in this scenario. Despite this being a tantalising hypothesis, Kolmogorov-Smirnov tests applied to the distributions for JClass 3 and 4, 3 and 5 and 4 and 5 galaxies yield p-values of 6.75, 13.5 and 7.06 per cent respectively, meaning we have a low confidence that a difference actually exists between the three distributions.

%--------------------------------------------------------------------
%--------------------------------------------------------------------

\begin{figure}
\includegraphics[width=\columnwidth]{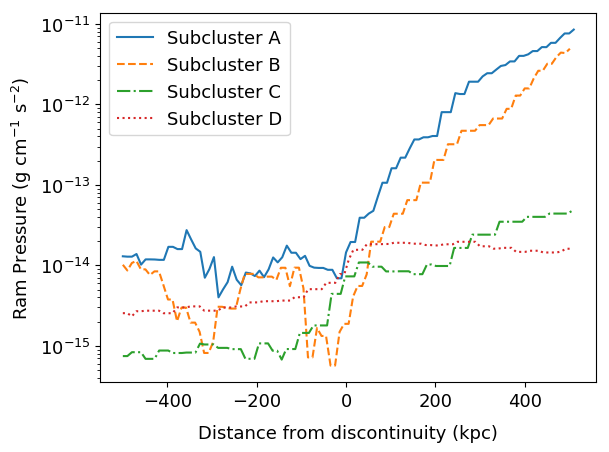}
\caption[]{Ram pressure variation along the four black lines shown in Fig.~\ref{fig:rampressures}, which are all perpendicular to their respective ram pressure boundaries. This illustrates the ram pressure increase which takes place in those boundaries, which can be of a factor of up to 1000.}
\label{fig:discontinuity}
\end{figure}

As a more quantitative illustration of the ram pressure variations along the ram pressure boundaries we report, we show in Fig.~\ref{fig:discontinuity} ram pressure profiles along the four black lines in Fig.~\ref{fig:rampressures}, which were chosen arbitrarily for the sake of illustration. Before the boundaries are crossed, the ram pressure profiles feature some noise, but overall they remain somewhat constant. After the boundary is crossed, an increase of a factor of 10 -- 1000 (depending on the cluster) takes place in the ram pressure within a few hundred kpc. The two largest subclusters (A and B) feature ram pressure increments larger than that of the remaining two (C and D), mainly due to their proximity to each other.

%--------------------------------------------------------------------
%--------------------------------------------------------------------

\begin{figure}
\includegraphics[width=\columnwidth]{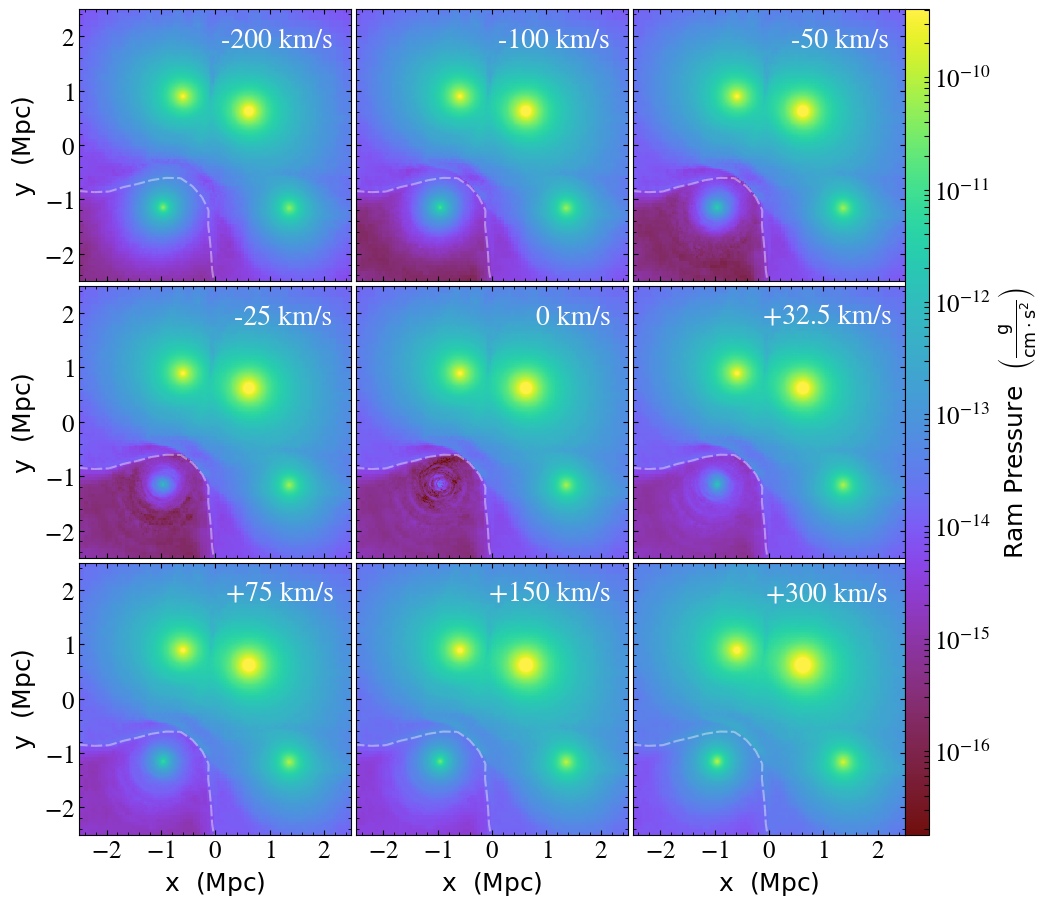}
\caption[]{Effect of peculiar velocities relative to the parent galaxy cluster. Each panel is a midplane ram pressure slice in a reference frame defined by the average velocity of the A902 subcluster plus the indicated velocity, added in the same direction. A ram pressure increase in the boundary we report is pronounced for velocities within 100 km/s of the average cluster velocity.}
\label{fig:differentspeeds}
\end{figure}

An initial hypothesis of the analysis so far was to calculate ram pressure in the reference frames of each subcluster, as an approximation for the velocities of its member galaxies. But the galaxies in reality feature peculiar velocities relative to their parent clusters. In Fig.~\ref{fig:differentspeeds} we show the same map as in Fig.~\ref{fig:rampressures} for subcluster C for reference frames with different velocities added in the same direction as that of the cluster average. We have chosen this subcluster for the sake of illustration because it is the one with the simplest ram pressure boundary. We find that the ram pressure boundary is only pronounced for velocities within roughly 100 km/s of the average cluster velocity; beyond that, the boundary fades away and the ram pressure intensity becomes correlated with the gas density at each location. This adds up to our picture so far in the following manner: our scenario should involve galaxies moving at relatively low speed relative to their parent subcluster, perhaps close to their apocentric passage. Those galaxies are still moving at high speed in the reference frame of the system as a whole, allowing them to cross the ram pressure boundary within a short timescale, of $\sim$100 Myr, and then become jellyfishes after that. 

%--------------------------------------------------------------------
%--------------------------------------------------------------------

One could also wonder whether our results would be different for off-plane ram pressure slices, i.e. planes parallel to the plane of the centres of the 4 subclusters, but at a certain height -- so far we have limited ourselves to a mid-plane slice. We have verified that the locations of the ram pressure boundaries are very close to that in the mid-plane slice for heights of up to $\sim$500 kpc; the main difference is that the off-plane densities are lower, making the off-plane ram pressure increments also smaller.

%--------------------------------------------------------------------
%--------------------------------------------------------------------

%%%%%%%%%%%%%%%%%%%%%%%%%%%%%%%%%%%%%%%%%%%%%%%%%%%
\section{Discussion and summary} \label{sec:discussion}

% Summary of our simulation
We have employed in our analyses a tailored simulation of the A901/2 system. Even though it consists of four subclusters, we were able to reach a satisfactory model. Dedicated simulations of galaxy cluster mergers involving three or more objects have not often been attempted -- an example is \citet{Bruggen2012}. In a general situation, the large number of degrees of freedom in the initial conditions would render the exploration of the parameter space nearly impracticable. However, in the particular case of A901/2, this approach was feasible due to some simplifying assumptions that were adopted --  apart from the usual setup of such idealised simulations (initially spherical objects, hydrostatic equilibrium, absence of small-scale substructures, etc). Regarding the dynamics of the system, our assumption was that the four objects are incoming for their first approach. This seems to be justifiable, given that they currently display no noticeable large-scale disturbances in their morphologies. Furthermore, we assumed for simplicity that the orbits are all on the plane of the sky. In our best model, the relative separations between the four subclusters are recovered. More importantly, the diffuse X-ray emissions are also quite well reproduced. This suggests that, in spite of the simplifying assumptions adopted, the gas properties in the simulation must be realistic within a good approximation.

From comparing the gas conditions in this model to the locations of the jellyfish galaxies in our adopted sample, we have inferred a possible mechanism for the triggering of jellyfish morphologies in this interaction environment, which is that the galaxies become jellyfishes when they cross a boundary within its parent subcluster where a significant increase in the ram pressure takes place, due to gas originally from the subcluster and gas from the remainder of the system meeting. A ram pressure increment of a factor of up to 1000 takes place on a scale of $\sim$100 kpc, while the galaxies are expected to be moving at speeds greater than 1000 km/s, implying a timescale of $\sim$100 Myr to cross the boundaries. We believe that this combination of a large increment in the ram pressure and a relatively short timescale to cross the boundaries makes it reasonable to conjecture that those boundaries could markedly affect the evolution of the gas content within the galaxies and turn them into jellyfishes. One caveat which should be pointed out is that this mechanism does not necessarily apply to all of the jellyfishes in the system. Indeed, in each subcluster a subset of galaxies far away from the respective boundary exists, which could have had their jellyfish morphologies triggered by a mechanism unassociated to the merger altogether. Still, we find significant differences in the distributions of distance to nearest boundary between the jellyfish sample and both a random sample and the STAGES sample of member galaxies in the system. This suggests that it is reasonable to assume that a significant fraction of the jellyfishes are indeed being generated by the aforementioned mechanism.

Another caveat is that the exact locations of the ram pressure boundaries in the real system could be different from the ones we report due to a variety of factors -- for instance, the exact positioning of clusters in the line of sight, deviations from our adopted initial density profiles and deviations from hydrostatic equilibrium in the clusters prior to the merger are all factors which could lead to different ram pressure distributions. Indeed, regarding deviations from hydrostatic equilibrium, cosmological simulations have pointed out that random bulk motions are relevant in the ram pressure stripping of galaxies in isolated galaxy clusters \citep{2008ApJ...684L...9T}. One illustrative implication of this is the following: if, for some reason, the boundary for subcluster C shown in Fig.~\ref{fig:rampressures} is in reality located some 100 kpc lower than what we find, then it would be located right on top of a concentration of about 8 jellyfish galaxies, which are all behind the boundary in our model.

% The mechanism is general
Although this entire work is devoted to the particular case of the A901/2 system, we expect the ram pressure boundaries we report inside each subcluster to be present in all galaxy cluster mergers which are still in the early stages of the collision, before their centres have crossed each other. In this way, a similar analysis to the one presented here could be carried out for other observed galaxy clusters, in order to probe the universality of the mechanism. Perhaps the biggest difficulty in this is finding jellyfish galaxies in clusters in the first place -- they are rare and their identification is very dependent on visual inspection.  
%it was only possible to obtain our jellyfish sample because the A901/2 system has been imaged with the HST telescope; current ground telescopes would not typically have the necessary angular resolution to allow such classification to be made, except for nearby clusters.
Previous observational work on jellyfish galaxies, such as \citet{2016MNRAS.455.2994M} and \citet{owers12}, have hinted that such galaxies could actually be preferentially found in galaxy cluster merger systems; our findings are consistent with interacting galaxy clusters being a favourable environment for searching for such galaxies.

% Summary of the our finding regarding the jellyfishes
The summary of this paper is as follows. We have developed a hydrodynamic model for the A901/2 system using a multi-cluster merger simulation, consistent with their positions relative to each other, their masses and their X-ray emissions. This model was used to correlate the gas conditions in the system with the locations of the jellyfish galaxy candidates in it identified by \mbox{\citet{roman-oliveira18}}. We have found that at each subcluster, a boundary exists where gas moving along the cluster and gas from the remainder of the system meet; in those boundaries, an increment of a factor of 10 -- 1000 in the ram pressure takes place within a few hundred kpc, due to a large increment in the diffuse gas velocity. More importantly, we have found that jellyfish galaxies in the system seem to be preferentially located near those boundaries, which could mean that the crossing of those boundaries is the mechanism behind the formation of jellyfishes at those locations. We propose that this mechanism could be common in galaxy cluster mergers which are at the beginning stages of their encounter, possibly making those environments particularly favorable for searching for jellyfish galaxies. This is the first theoretical treatment of ram pressure stripping in the environment of galaxy cluster mergers which has been presented in the literature.

%%%%%%%%%%%%%%%%%%%%%%%%%%%%%%%%%%%%%%%%%%%%%%%%%%%
\section*{Acknowledgements}

RR thanks the S\~ao Paulo Research Foundation, FAPESP, for the financial support (grant 15/13141-7). This work has made use of the computing facilities of the Laboratory of Astroinformatics (IAG/USP, NAT/Unicsul), whose purchase was made possible by the Brazilian agency FAPESP (grant 2009/54006-4) and the INCT-A. Simulations were carried out in part at the Centro de Computa\c c\~ao Cient\'ifica e Tecnol\'ogica (UTFPR). This study was financed in part by the \textit{Coordena\c{c}\~{a}o de Aperfei\c{c}oamento de Pessoal de N\'ivel Superior - Brasil} (CAPES) - Finance Code 001. 
%LD acknowledges support from the Brazilian agency CAPES.
ACS and FRO acknowledge funding from the Brazilian agencies \textit {Conselho Nacional de Desenvolvimento Cient\'ifico e Tecnol\'ogico} (CNPq) and the Rio Grande do Sul Research Foundation (FAPERGS) through grants PIBIC-CNPq, CNPq-403580/2016-1, CNPq-310845/2015-7, PqG/FAPERGS-17/2551-0001, PROBIC/FAPERGS and L'Or\'eal UNESCO ABC \textit{Para Mulheres na Ci\^encia}. 
GBLN thank financial support from CNPq and FAPESP (2018/17543-0). BRP acknowledges financial support from the Spanish Ministry of Economy and Competitiveness through grants ESP2015-68964 and ESP2017-83197. We thank the anonymous reviewer for the feedback, which helped make the presentation of our results more clear.

%%%%%%%%%%%%%%%%%%%%%%%%%%%%%%%%%%%%%%%%%%%%%%%%%%%
\bibliographystyle{mnras.bst}
\bibliography{A901.bib}

%%%%%%%%%%%%%%%%%%%%%%%%%%%%%%%%%%%%%%%%%%%%%%%%%%%
\appendix
\section{Code comparison} \label{sec:codecomparison}

%--------------------------------------------------------------------
\begin{figure}
\includegraphics[width=\columnwidth]{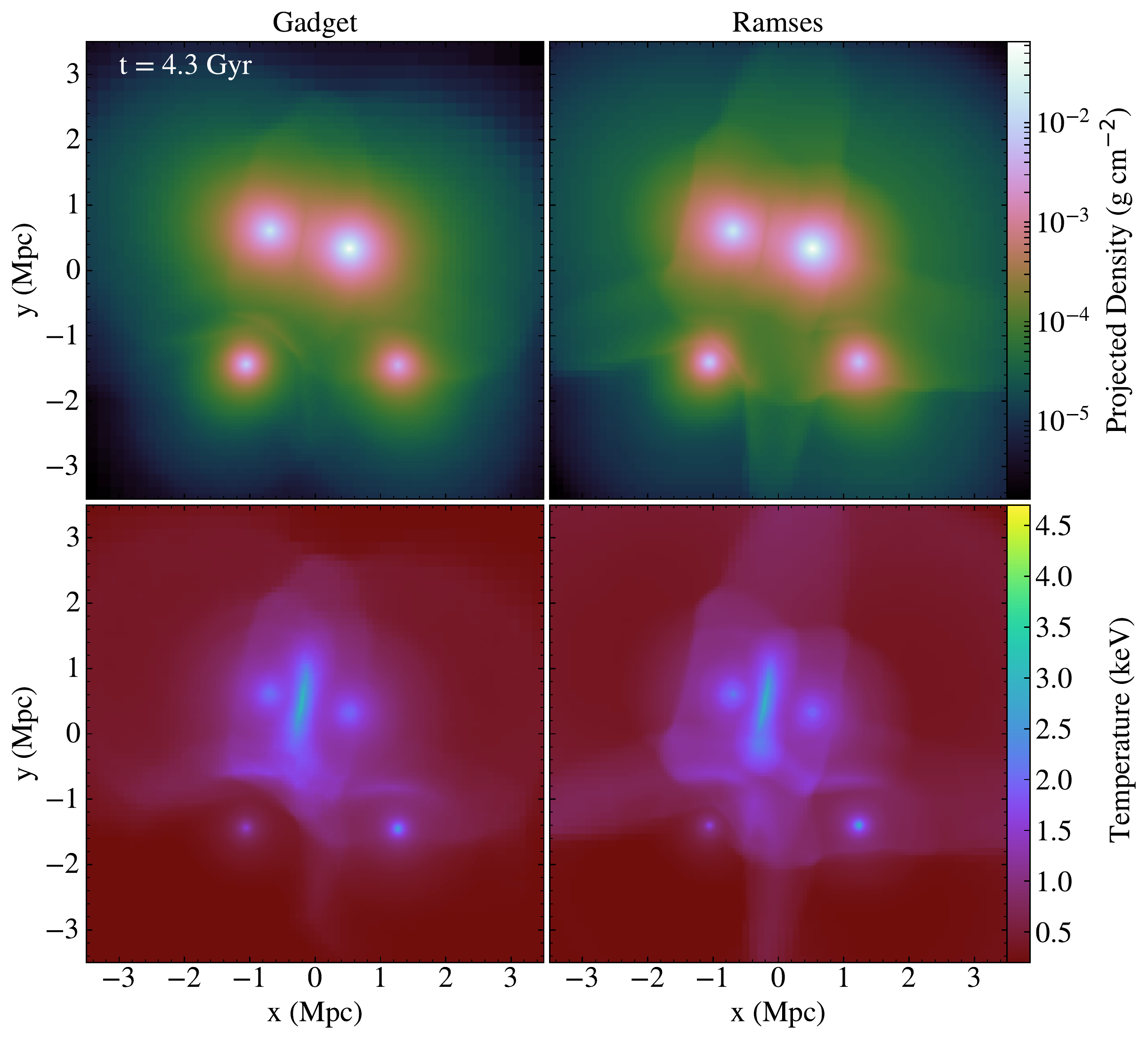}
\caption[]{Comparison between {\sc gadget-2} (left) and {\sc ramses} (right) results, showing gas density maps (upper panels) and temperature maps (lower panels), for $t=4.3$\,Gyr.}
\label{fig:ramses43}
\end{figure}
%--------------------------------------------------------------------

To compare the hydrodynamic evolution results obtained from the simulation using the code {\sc gadget-2}, new simulations were performed with the {\sc ramses} code. Both methods attempt to solve the fluid equations, but in very different ways:  {\sc gadget-2} is a cosmological simulation code based on smoothed particle hydrodynamics (SPH) technique that computes gravitational forces with a hierarchical tree algorithm \citep{Springel2005}, whereas {\sc ramses} is based on an Adaptive Mesh Refinement (AMR) technique, with a tree based data structure allowing recursive grid refinements on a cell-by-cell basis \citep{Teyssier2002}. Both runs used the same initial conditions, and with comparable resolution (the minimum and maximum refinement levels defined on {\sc ramses} were 6 and 12 respectively). A mass-based refinement criterion was employed on the {\sc ramses} run, in order to ensure that the discretisation was equivalent to that of a particle-based code like {\sc gadget-2}.

The analysis and of the output was done using the {\sc yt} analysis code \citep{Turk2011}, so it was possible to create Fig.~\ref{fig:ramses43}, showing the projected density and temperature maps for the instant of time $t=4.3$\,Gyr. When comparing the two codes, it can be noted that the final coordinates of the objects are in good agreement, that is, the global morphology is quite similar (with the possible exception of small-scale details). Similarly, the ranges of temperature are comparable, even in regions of intense variations. This overall agreement is consistent with other studies on the comparison between AMR and SPH simulations as shown in \cite{OShea2005}, \cite{Hubber2013} and \cite{Kim2014}.

\label{lastpage}

\end{document}